\pdfoutput=1
\documentclass[aps,pre,twocolumn]{revtex4} 
\usepackage{hyperref,natbib,doi}
\usepackage{graphicx}
\usepackage{amsmath}
\usepackage{dcolumn}
\usepackage[utf8]{inputenc}
\usepackage{amssymb}
\begin{document}
\title{Differences in 1D electron plasma wake field acceleration in MeV versus GeV and
linear versus blowout regimes}
\author{David Tsiklauri}
\affiliation{School of Physics and Astronomy, Queen Mary University of London, London, E1 4NS, United Kingdom}
\begin{abstract}
In some laboratory and most astrophysical situations
plasma wake-field acceleration of electrons is one dimensional, i.e.
variation transverse to the beam's motion can be ignored. Thus,
one dimensional (1D), particle-in-cell (PIC), fully electromagnetic simulations of 
electron plasma wake field acceleration are conducted in order to study the 
differences in electron plasma wake field acceleration in MeV versus GeV and
linear versus blowout regimes.
First, we show that caution needs to be taken when using
fluid simulations, as PIC simulations prove that an approximation for an
electron bunch not to evolve in time for few hundred plasma periods 
only applies when it is sufficiently relativistic.
This conclusion is true irrespective of the plasma temperature.
We find that in the linear regime and GeV energies,
the accelerating electric field generated by the
plasma wake is similar to the linear and MeV
regime. However, because GeV energy driving bunch
stays intact for much longer time, the final acceleration energies
are much larger in the GeV energies case.
In the GeV energy range and blowout regime
the wake's accelerating electric field is much
larger in amplitude compared to the linear case and also
plasma wake geometrical size is much larger.
Thus, the correct positioning of the trailing bunch is needed to
achieve the efficient acceleration.
For the considered case,  optimally there should be approximately $(90-100) c/\omega_{pe}$
distance between trailing and driving electron bunches in the GeV blowout regime.
\end{abstract}

\maketitle

\section{Introduction}

The plasma acceleration 
based on laser wake field acceleration 
stems from a paper by Tajima and Dawson \citep{td79}. 
When a laser is injected in plasma,
 a ponderomotive force arises
from the nonlinear Lorentz force $v \times B/c$, which causes the polarization of electrons in
the plasma in the longitudinal direction, even though the electric field of the laser is
in the transverse direction. This polarization $E_p = m_e \omega_p c a_0/e$ yields the electrostatic
field in the longitudinal direction with the same order of magnitude.
Here the normalized vector potential of the laser is $a_0 = e E_0/m_e \omega_0 c$ and
$E_0$, $\omega_0$ are the electric field and frequency of the laser.
The relativistically intense laser makes the amplitude of the wake-field $E_p$
relativistically intense, i.e. $a_p = eE_p/m_e \omega_p c \gg 1$ \citep{tajima17}. 
The plasma wave breaking occurs at $a_0 \approx 1$.
The experimental implementation of the plasma wake field was done
by Joshi \citep{joshi}.
There are two main possibilities for creation of the plasma wake:
a laser or an electron bunch. 
The former is referred to as laser wake field acceleration (LWFA)
and the latter as plasma wake field acceleration (PWFA).
In early experiments, injected electrons of a few
MeV have been accelerated by GV/m electric fields using
 LWFA \citep{1998PhRvL..81..995A}.
  In these experiments, because the bunch length of
the injected electrons was much longer than the plasma
wavelength, only a small fraction of the injected
electrons were accelerated. In turn, this results in a poor
electron beam quality \citep{doi:10.1063/1.3695389}.
More recent experimental devices with {\it compact} electron beams 
show accelerating gradients several 
orders of magnitude better (10s of GeV m$^{-1}$) than current 
RF-based conventional particle accelerators (10s of MeV m$^{-1}$).
A significant progress in PWFA has been made recently 
both in experiment and theory \cite{l14,lau15,farinella16,me2016,bera16}.

Good progress has been made in applying PWFA concepts to
astrophysical plasmas. Including supermassive black hole \cite{Ebisuzaki20149,
Ebisuzaki2014}, and solar atmosphere \cite{me2017} contexts.

Section 2 presents the model and results. In Section 3 we list the main findings.

\section{The model and results}

\begin{figure*}
\includegraphics[width=\textwidth]{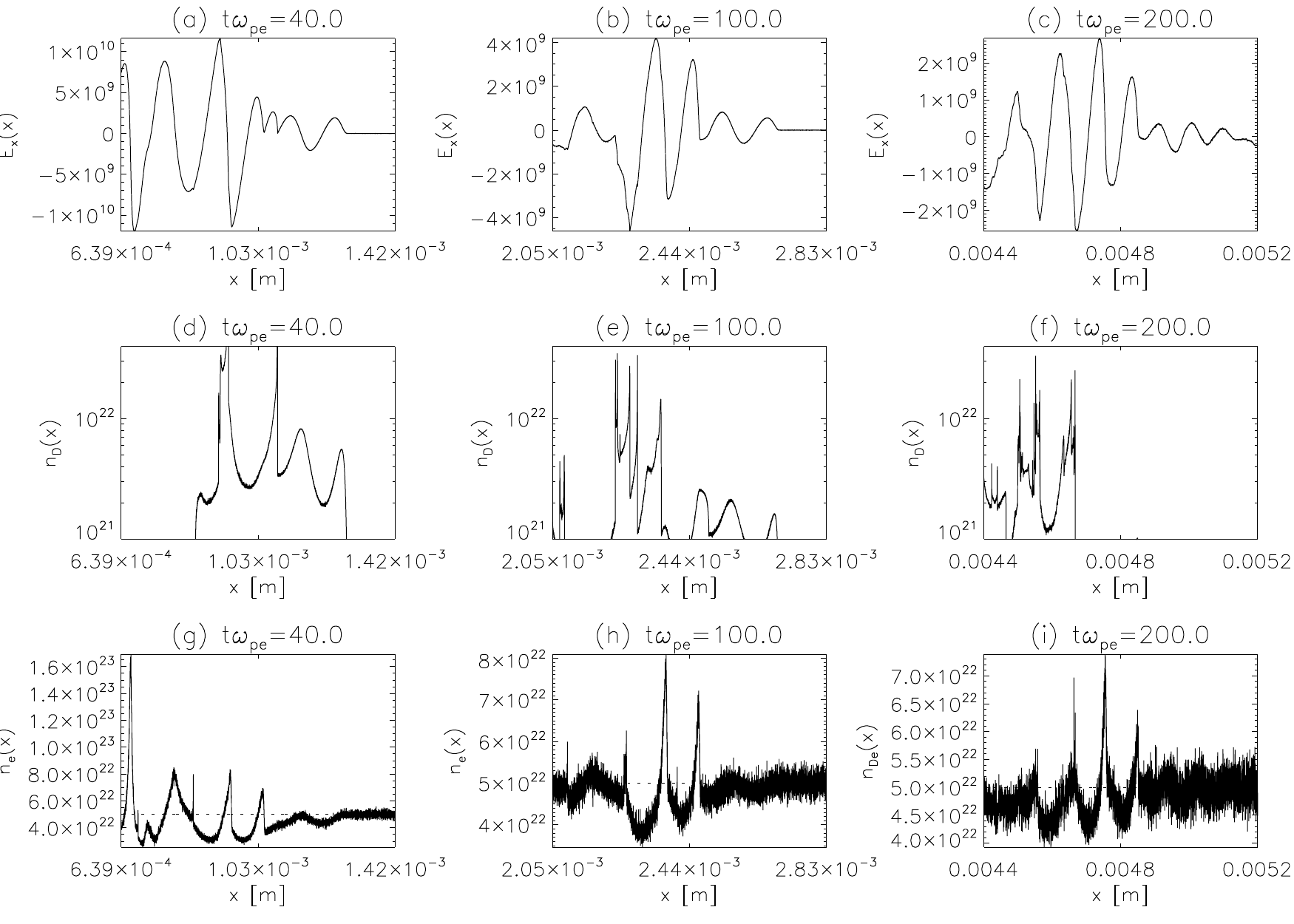}
\caption{(a-c) electric field x-component 
at different time instants
corresponding to 1/5th, half and the final simulations times.
(d-f) log normal plot of driving
electron bunch number density at the same times.
(g-i) Solid line is background electron number density, in units of m$^{-3}$,
 at the same times. Dashed line is initial ($t=0$) background electron number density of
 $5\times 10^{22}$ m$^{-3}$, to guide the eye.
The fields are quoted in $V/m$ and time at the top of each panel is
in $ \omega_{pe}$. The data is for Run 1. See text and table \ref{tab1} for details.
Note that x-coordinate is different in each raw because we use a window which follows the 
bunch with
speed $v_b$.}
\label{fig1}
\end{figure*}

\begin{figure*}
\includegraphics[width=\textwidth]{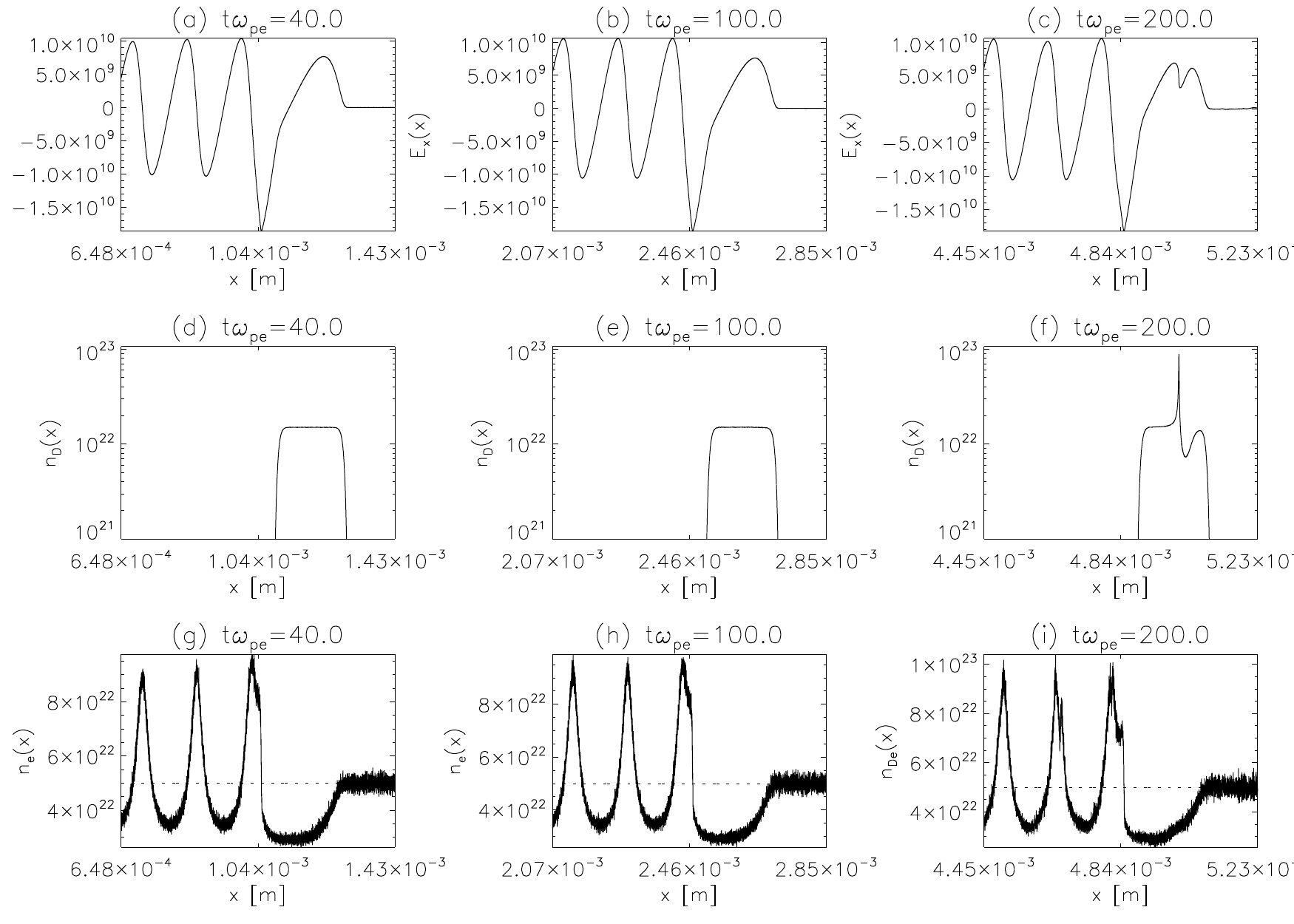}
\caption{As in Fig.\ref{fig1} but for Run 2.}
\label{fig2}
\end{figure*}

\begin{figure*}
\includegraphics[width=\textwidth]{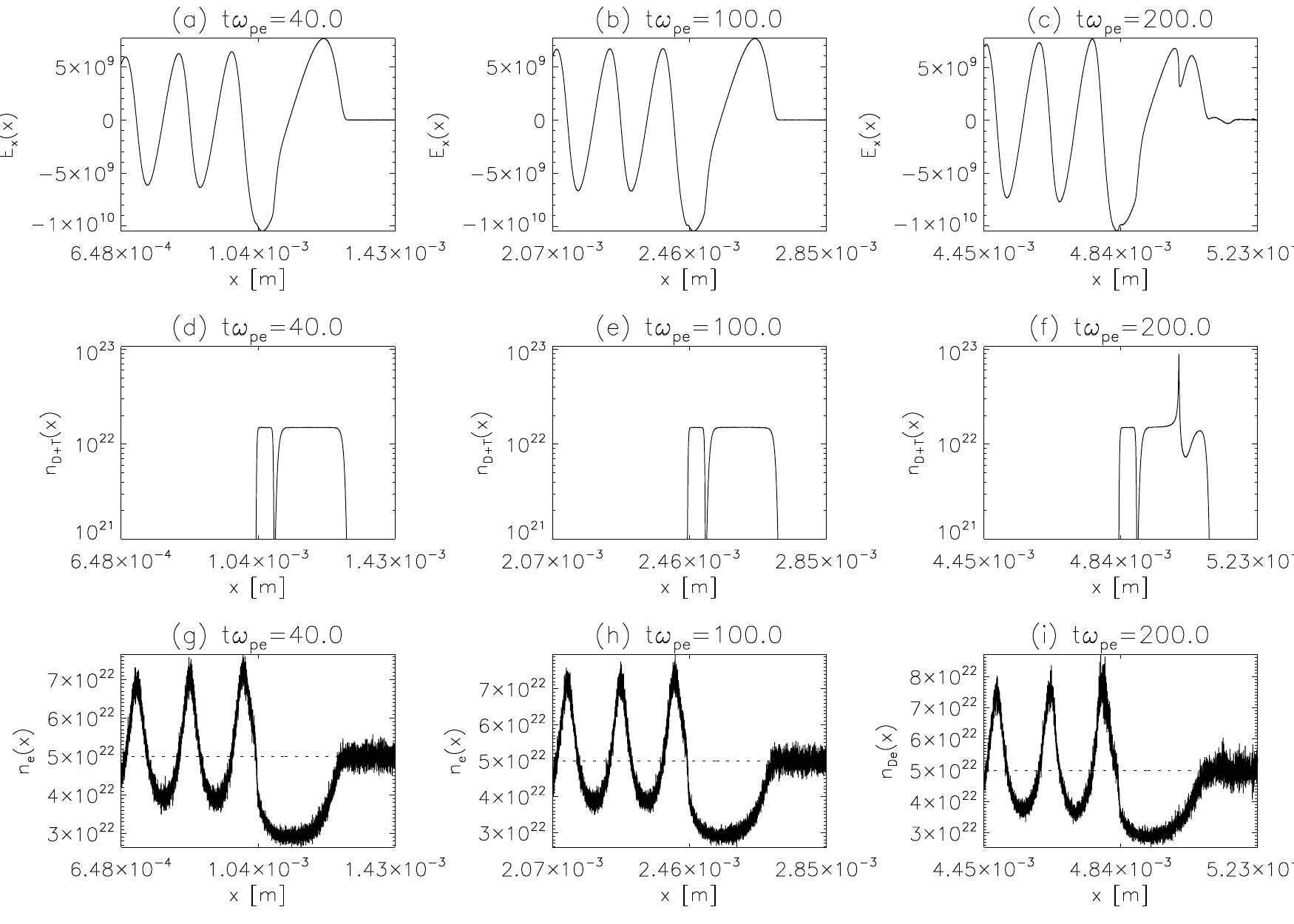}
\caption{ As in Fig.\ref{fig1} but for Run 3. Note that panels (d-f)
now also include trailing electron bunch contribution.}
\label{fig3}
\end{figure*}

\begin{figure*}
\includegraphics[width=\textwidth]{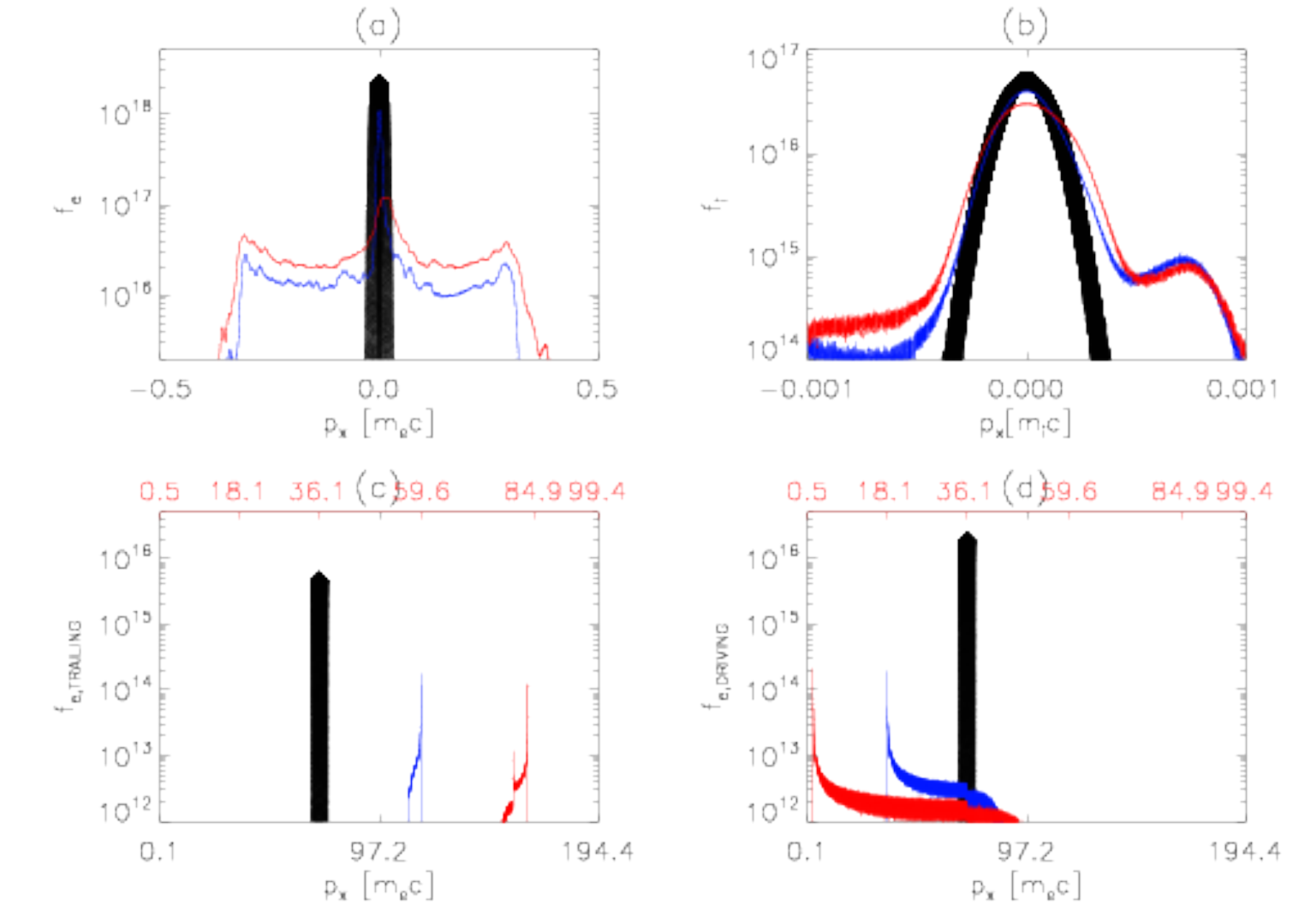}
\caption{Background electron (a), ion (b), trailing (c) and 
driving (d) electron bunch
distribution functions shown at different times in different colors:
open diamonds correspond to $t=0$, while blue and red curves to the 
half and the final simulations times, respectively. On
x-axis the momenta are quoted in the units of relevant species mass times
speed of light i.e. $[m_e c]$ or $[m_i c]$. 
At the top of panels (c) and (d), to guide the eye,  the energies is quoted in MeV with red numbers.
The data is for Run 3.}
\label{fig4}
\end{figure*}

\begin{figure} 
\includegraphics[width=\columnwidth]{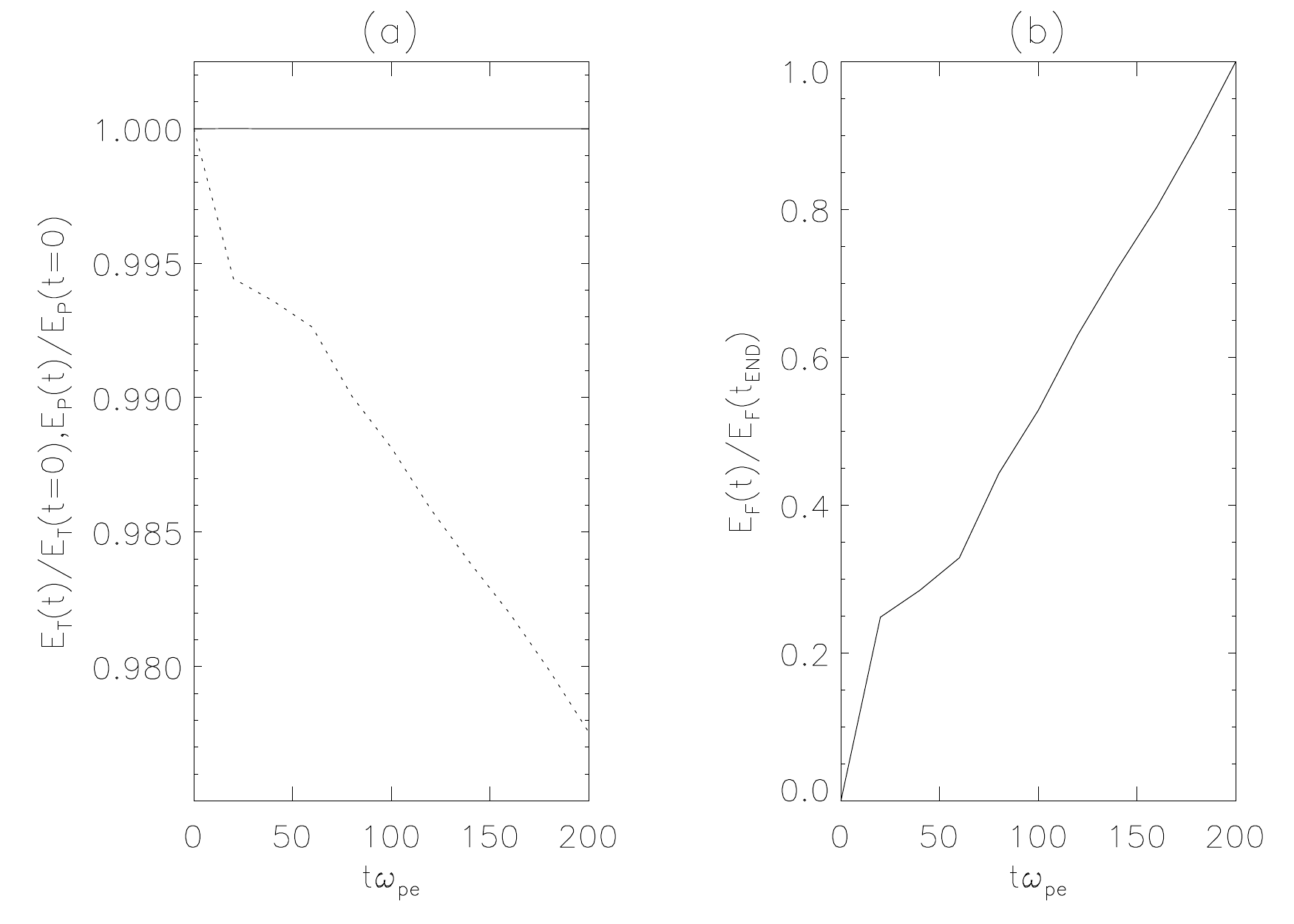}
\caption{Panel (a) solid and dashed curves respectively
are the total (particles plus EM fields) and only
particle energies, normalized on their initial values. 
Panel (b) shows
EM field energy, normalized on its final simulation time
value. 
The data is for Run 3.}
\label{fig5}
\end{figure}

\begin{figure*}
\includegraphics[width=\textwidth]{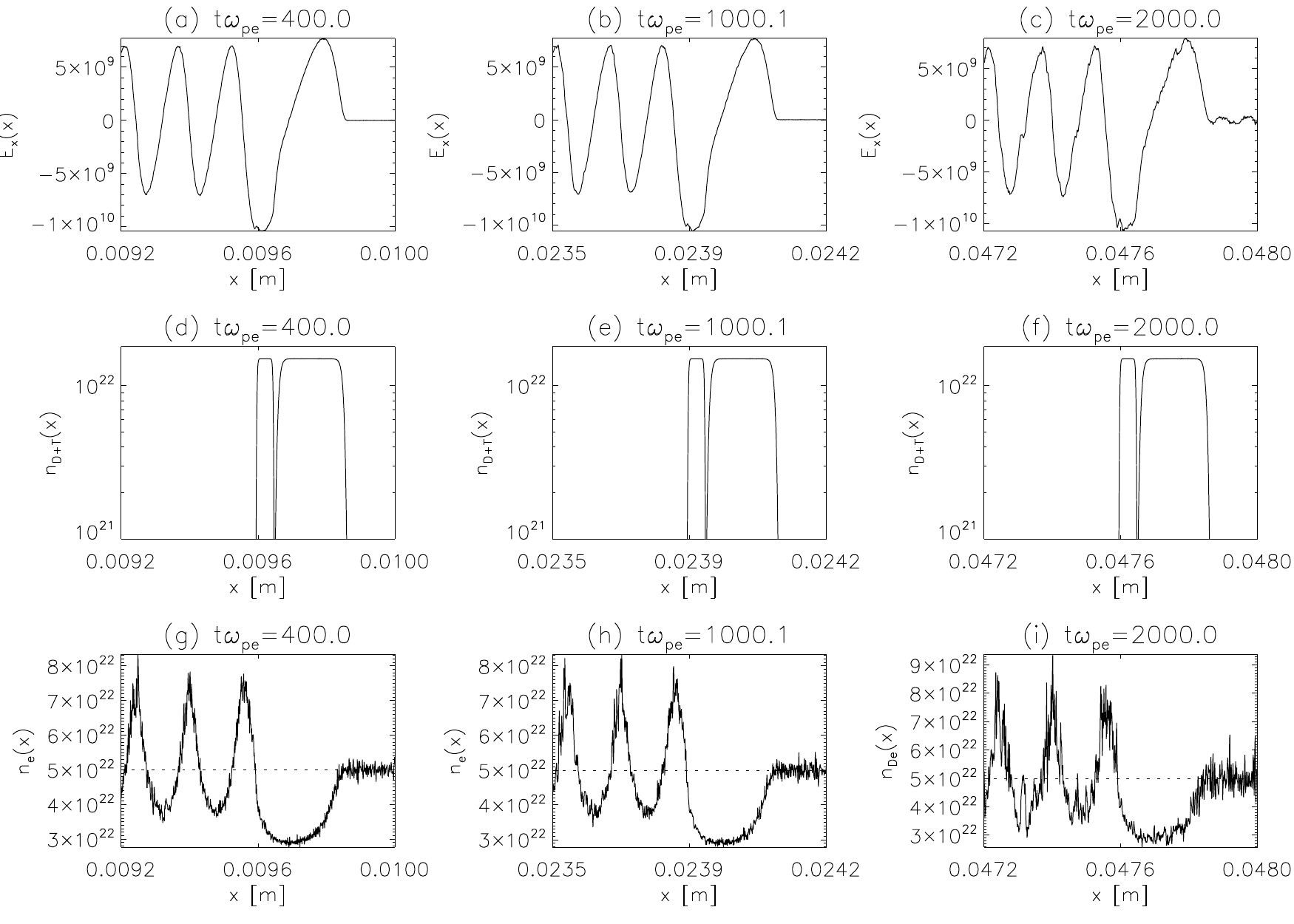}
\caption{ As in Fig.\ref{fig3} but for Run 4.}
\label{fig6}
\end{figure*}

\begin{figure*}
\includegraphics[width=\textwidth]{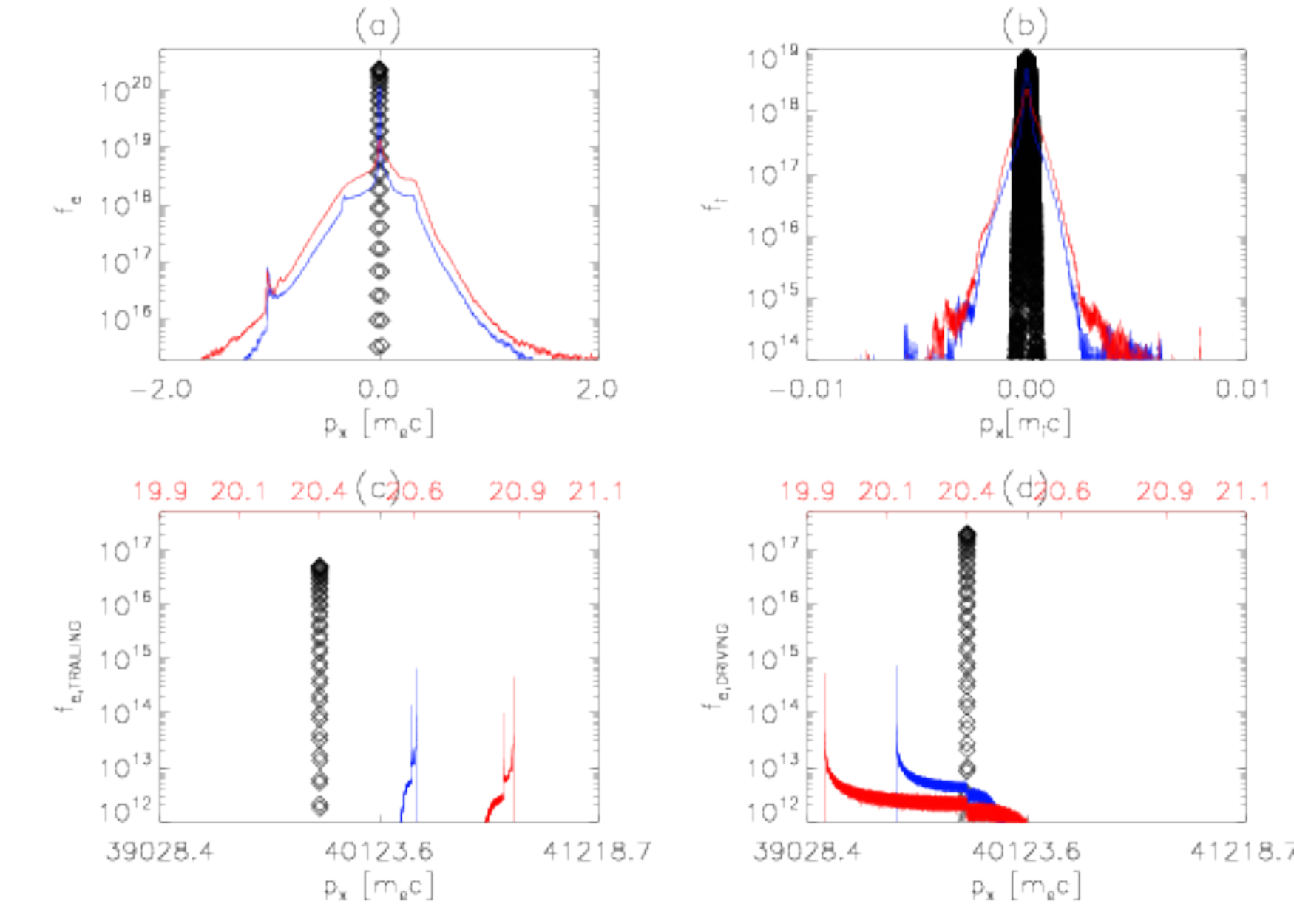}
\caption{As in Fig.\ref{fig4} but for Run 4. Note that energies 
in the red, top scale in panels (c) and (d) are now
in GeV.}
\label{fig7}
\end{figure*}

\begin{figure*}
\includegraphics[width=\textwidth]{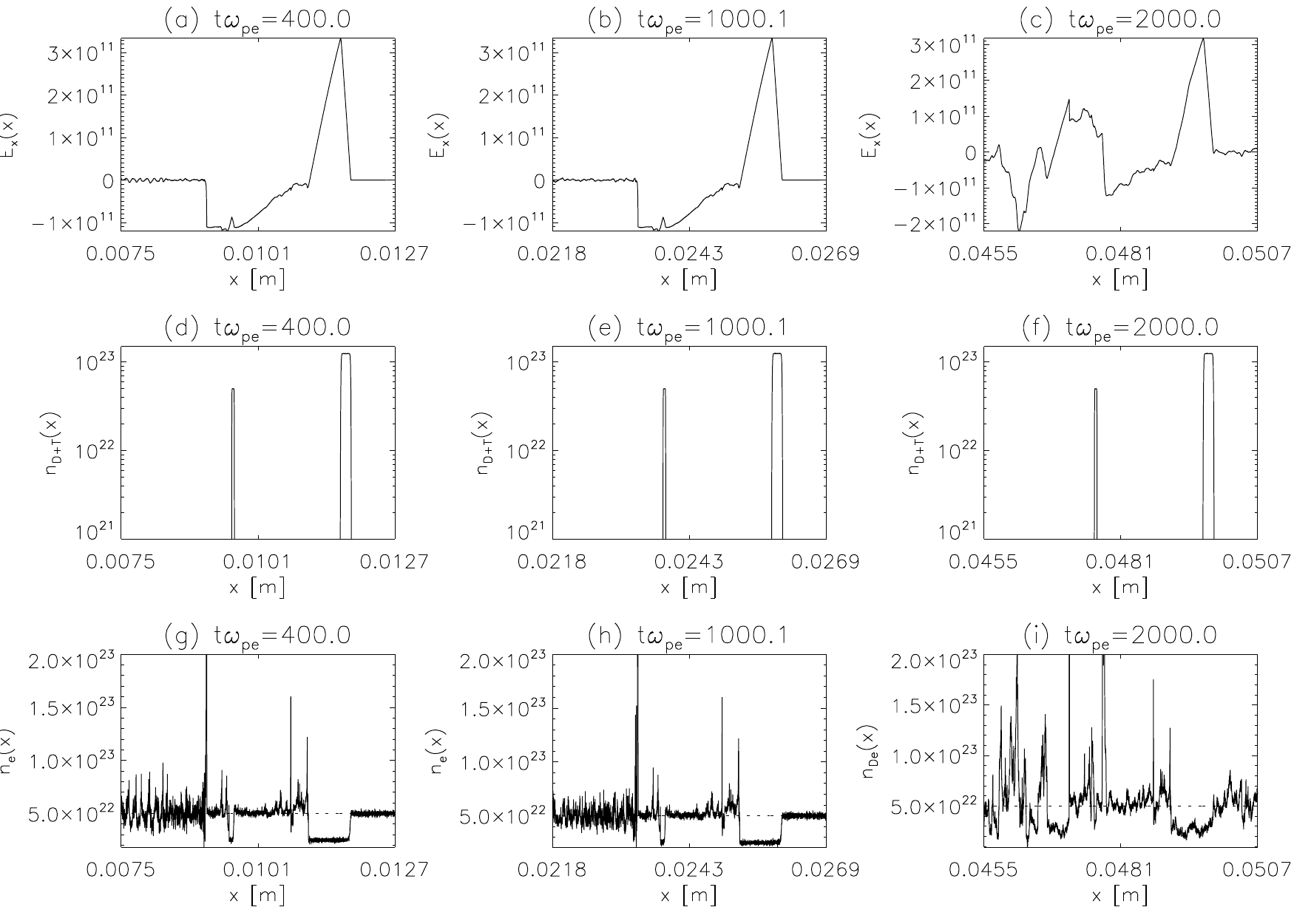}
\caption{ As in Fig.\ref{fig3} but for Run 5.}
\label{fig8}
\end{figure*}

\begin{figure*}
\includegraphics[width=\textwidth]{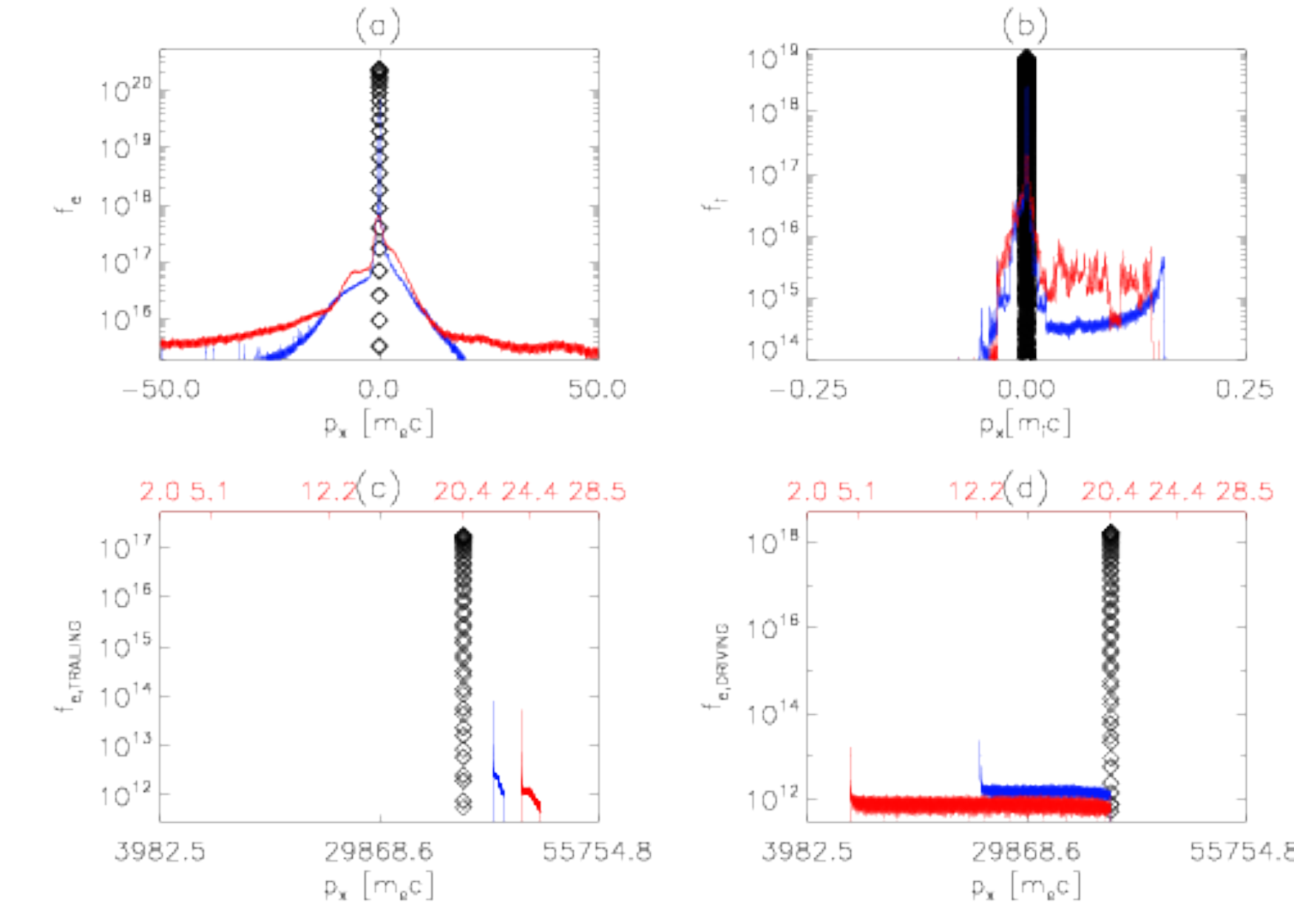}
\caption{As in Fig.\ref{fig4} but for Run 5. Again, energies in the 
red, top scale in panels (c) and (d) are in GeV.}
\label{fig9}
\end{figure*}

\begin{figure*}
\includegraphics[width=\textwidth]{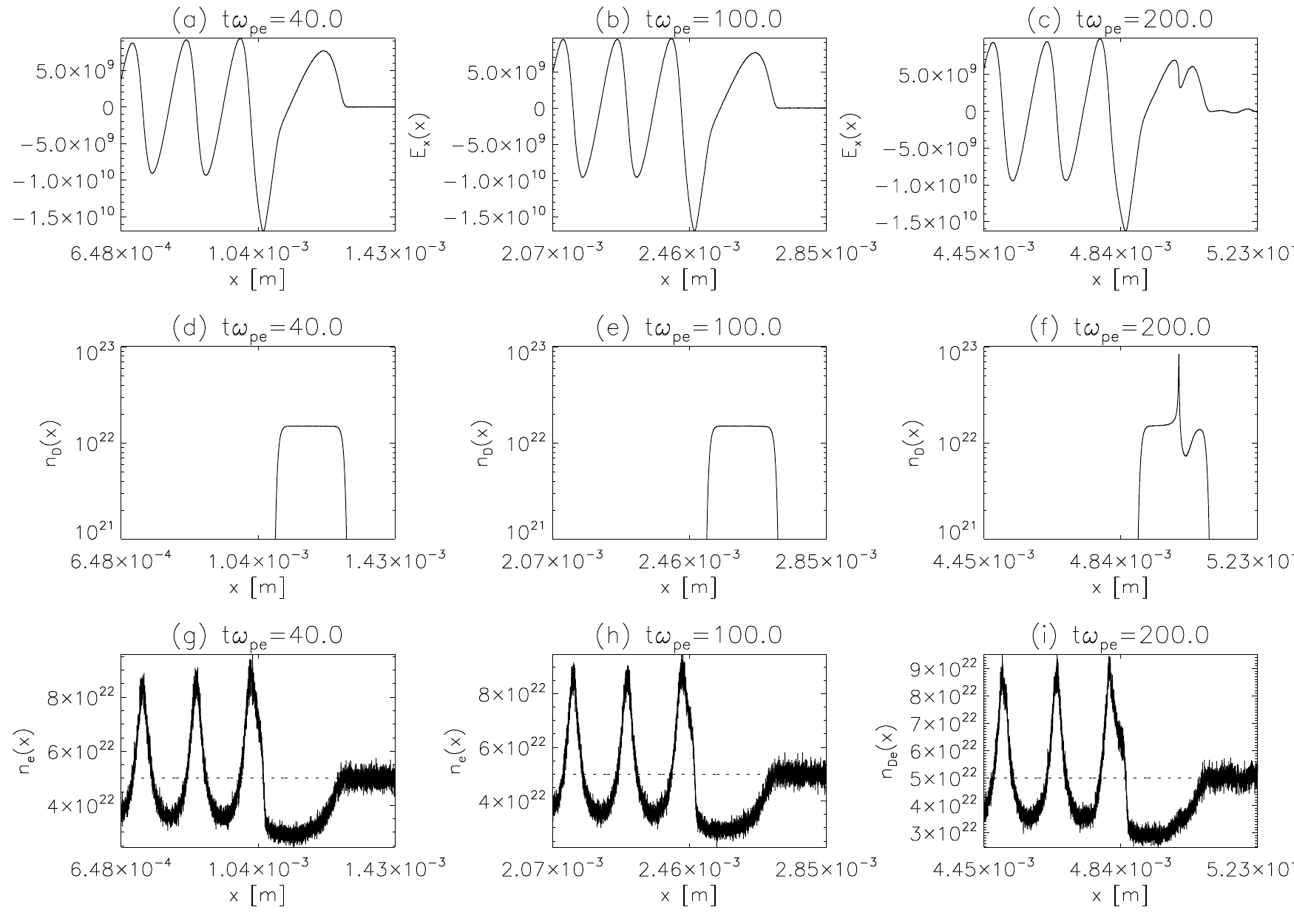}
\caption{As in Fig.\ref{fig1} but for numerical run similar to Run 2
now with 100 times hotter temperature and 10 times shorter domain length.}
\label{fig10}
\end{figure*}

\begin{figure*}
\includegraphics[width=\textwidth]{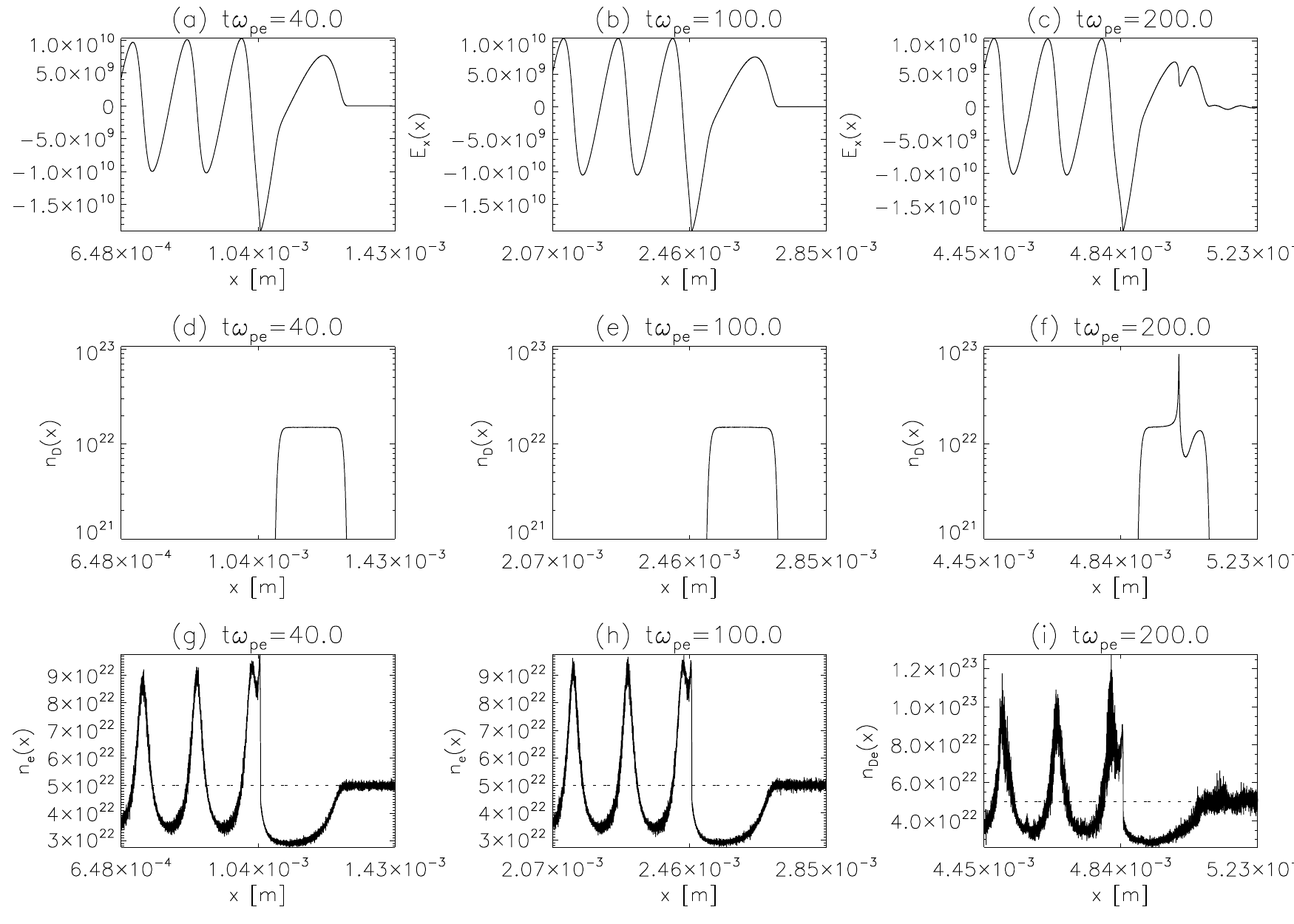}
\caption{As in Fig.\ref{fig1} but for numerical run similar to Run 2
now with 100 times cooler temperature and 10 times longer domain length.}
\label{fig11}
\end{figure*}

We used EPOCH, a
fully electromagnetic (EM), relativistic PIC code \cite{a15} for the simulation.
EPOCH is available for download from
\url{https://cfsa-pmw.warwick.ac.uk}.
The mass ratio in all runs is
$m_i/m_e=1836.153$ and boundary conditions are periodic.
Choice of boundary conditions is not
important here because simulation domain is long enough,
such that the electron bunches never reach the boundary.

The simulations domain is split into
$n_x =65280$
grid cells in x-direction.
We fix grid size $\Delta$ as Debye length ($\lambda_D$) times appropriate factor ($f$), i.e.
$\Delta= f \lambda_D $.
Here $\lambda_D = v_{th,e}/\omega_{pe}$ denotes the Debye length
with 
$v_{th,e}=\sqrt{k_B T/m_e}$ being electron thermal
speed and $\omega_{pe}$ electron plasma frequency.
In the plasma wake field acceleration
the relevant spatial scale is electron inertial length,
$c/\omega_{pe}$. 
We vary factor $f$ such that: 
(i) in runs 1 -- 3 $f=1$ and $c/\omega_{pe}$ is
resolved with 243 grid points, i.e.
$(c/\omega_{pe})/\Delta=243.5$, where again
$\Delta=f  \lambda_D$ is the grid size;
(ii) in runs 4 and 5 $f=10$ and $c/\omega_{pe}$ is
resolved with also 24 grid points, i.e.
$(c/\omega_{pe})/\Delta=24.35$.
This choice provides a good
resolution
as the energy error never exceeds $\approx 0.0004\%$.

\begin{table}[htbp]
\caption{Conducted numerical run details. 
$N_b$ stands for number of electron bunches.
"1" means presence only one (driving) bunch.
"2" means presence of two, driving and trailing bunches.
$E_i$ denotes bunch initial energy.
$t_{end}$ is simulation end time in units of $\omega_{\mathrm{pe}} $.
$E_{f,T}$  is final energy of the trailing bunch.
$A_D=n_D/n_0$  is ratio of driving bunch and background electron number densities.
$A_T=n_T/n_0$ is ratio of trailing bunch and background electron number densities.} 
\centering 
\begin{tabular}{c c c c c c c} 
\hline\hline 
Case & $N_b$ & $E_i$ & $t_{end}$  & $E_{f,T}$  & $A_D$ & $A_T$ \\ [0.5ex] 
\hline %
Run 1   & 1 & 3.6 MeV & 200  & N/A    & 0.3 & N/A\\ 
Run 2   & 1 & 36 MeV  & 200  & N/A    & 0.3 & N/A\\
Run 3   & 2 & 36 MeV  & 200  & 85 MeV & 0.3 & 0.3 \\
Run 4   & 2 & 20 GeV  & 2000 & 21 GeV & 0.3 & 0.3 \\
Run 5   & 2 & 20 GeV  & 2000 & 24 GeV & 2.5 & 1.0 \\ [1ex]  
\hline
\end{tabular}
\label{tab1} 
\end{table}

Different numerical runs are described in table \ref{tab1}.
At $t=0$, the driving and trailing electron  bunches have the number
densities as follows:
\begin{equation}
n_{D}(x)= A_D n_0  \exp\left[-\frac{(x-x_D c/\omega_{pe})^{16}}
{(4.0 c/\omega_{pe})^{16}} \right]  
\label{e1},
\end{equation}
\begin{equation}
n_{T}(x)= A_T n_0  \exp\left[-\frac{(x-4.5 c/\omega_{pe})^{16}}
{(c/\omega_{pe})^{16}} \right]  
\label{e2},
\end{equation}
where $x_D$ is location of the driving bunch in units of $c/\omega_{pe}$
and it has a length of $4.0 c/\omega_{pe}$, as in Figs. 1 and 2  from
\citet{bera16}, for easy inter-comparison.
$x_D = 10 c/\omega_{pe}$ in Runs 1--4 and $x_D = 94.5 c/\omega_{pe}$  in Run 5.
Trailing bunch that is absent in \citep{bera16} simulation is located at
 $4.5 c/\omega_{pe}$ and has a length of $c/\omega_{pe}$.
$A_D$ and $A_T$ are the bunch amplitudes in units of $n_0$.
If trailing bunch is present then both electron bunch initial momenta are set to
values shown in table \ref{tab1}.
For example for Runs 2 and 3 we set 
$p_x=p_0=\gamma m_e 0.9999c$ kg m s$^{-1}$ 
(note that $p_x/(m_e c)=70.7$, i.e. $\gamma=70.7$),
which corresponds to an initial
energy of $E_0=36.1$ MeV.
In the simulation with both bunches, there are four  plasma species present:
background electrons and ions, plus
driving and trailing bunches.
In the numerical runs there are 
$256$ particles per cell for each of the four species.
The numerical Runs 1--3 take about 3 hours on 96 cores using
Intel Xeon E5-2683V3 (Broadwell) processors
with 256GB of RAM and Mellanox ConnectX-4 EDR Infiniband Interconnect.
Runs 4 and 5 take 5 hours each on 192 cores of the same processors.
 
\subsection{Runs 1 and 2 -- importance of large $\gamma$-factor.}

In Run 1 we put to test fluid-simulation results of \citet{bera16}. The 
physical parameters are
as in their Figure 1 and are stated in table \ref{tab1}.

Fig.\ref{fig1} top row, panels (a-c), shows 
electric field x-component 
at different time instants
corresponding to 1/5th, half and the final simulations times.
We see that as the electron bunch moves in plasma it generates a wake and 
at early time electric field is of the order of $-10^{10}$ V/m,
but by the end simulation time is depletes to $-2 \times 10^{9}$ V/m.
This is understandable, because, as we see in 
Fig.\ref{fig1}(d-f) where we show log normal plots of driving
electron bunch number density at the same times,
the electron beam completely disintegrates. Commensurate plots in 
panels Fig.\ref{fig1}(g-i) of background electron number density
also show that the wake weakens. 
Fig.\ref{fig2} shows the results of Run 2 which has all the same parameters of
Run 1, except $v_b=0.9999 c$ ($E_0=36$ MeV), while in Run 1 $v_b=0.99 c$ ($E_0=3.6$ MeV).
We gather from panels Fig.\ref{fig2}(a-c) that the electric field of the wake
is not changing significantly (noticeably) in time.
Also, Fig.\ref{fig2}(d-f) shows that the electron bunches
stay intact and only start showing minor shape distortions in the
leading bunch by time $t \omega_{pe}=200$.
We see from from panels Fig.\ref{fig2}(g-i)  that electron number density
also is stable and resembles closely both analytical solutions
and fluid simulation results show in Fig. 1 of \citet{bera16}.
We do not overplot their analytical solution for brevity here, but
the closeness of match is obvious.
The main conclusion from Figs.\ref{fig1} and \ref{fig2} particularly from respective panels (d-f),
is that caution is needed when using
fluid simulations \citep{bera16}, as our PIC simulations prove that an approximation for an
electron bunch not to evolve in time for few hundred plasma periods 
only applies when it is sufficiently relativistic,
i.e. $v_b=0.9999 c$ and not mere $v_b=0.99 c$.

\subsection{Run 3 -- acceleration in the MeV energy range}

Now that we established that for  $v_b=0.9999 c$ the plasma wake is stable and also judging from
the structure and spatial extent of negative values of electric field  from Fig.\ref{fig2}, panels (a-c),
we perform new Run 3, now also adding a trailing bunch of length $c/\omega_{pe}$ and at position
prescribed by equation (2).
Such length and position deliberately coincides with
large negative dip seen in  Fig.\ref{fig2} and  Fig.\ref{fig3} panels (a-c).

According to \citet{bulanov16} there are two effects that work against an efficient electron acceleration:
(i) depletion of either driving laser pulse or electron bunch and
(ii) de-phasing of the trailing electron bunch from the negative electrostatic $E_x$ plasma
wake. 
Naturally only negative electric field can accelerate the electrons.
The positive one, on contrary, produces deceleration.
Both the electron slippage with respect to the accelerating phase of the wake and
the driving bunch/laser pulse energy depletion are both important. 
Comparing panels (a-c)  
to panels (d-f) in Fig.\ref{fig3} we see that trailing bunch is co-spatial with 
negative dip in $E_x$. 
This means that driving bunch will be decelerating,
while trailing bunch accelerating, because of sign of $E_x$.
Note that in the end of simulation time $t \omega_{pe}=200$ in panel (f)
the leading bunch starts to develop some rather localized spikes.
This means that the first effect (depletion) becomes into play.
We thus stop simulation at $t \omega_{pe}=200$.
From the panels (g)-(i) we gather that initially there was a substantial cavity
created in the background electrons.

In Fig.\ref{fig4} we plot the background electron (a), ion (b), trailing (c) and 
driving (d) electron bunch
distribution functions at different times in different colors:
open diamonds correspond to $t=0$, while blue and red curves to the 
half and the final simulations times, respectively. These data correspond to Run 3.
We gather from panel (a) that background electrons develop
broad peaks corresponding roughly to the momenta $p_x/(m_e c)\approx \pm 0.3$
is probably due to acceleration of trapped background electrons in positive
and negative peaks of $E_x$. In panel (b) we see that background ions
develop a beam with positive velocities corresponding
to momentum $p_x/(m_i c)\approx 0.0008$.
Panel (c) shows that by the end of simulation
 the trailing bunch gains energy to 85 MeV (red curve), starting from initial 36.1 MeV.
Panel (d) shows that by the end of simulation
 the driving bunch loses energy to 0.5 MeV (red curve), starting from initial 36.1 MeV.
This demonstrates that trailing electron bunch acceleration is
due to deceleration of driving bunch. The same conclusion 
follows from the dynamics of different kinds of energies in Fig.\ref{fig5}.

In  Fig.\ref{fig5} panel (a) solid and dashed curves respectively
are the total (particles plus EM fields) and only
particle energies, normalized on their initial values. 
Panel (b) shows
EM field energy, normalized on its final simulation time
value. Because at $t=0$ all EM fields are zero, hence initial 
EM field energy cannot be used for normalization. 
The data is for Run 3.
The total normalized energy 
stays constant and is approximately unity.
Its maximal deviation from unity is 0.000004 i.e. $0.0004 \%$
is due to numerical
heating and numerical dissipation (due to finite differencing).
The particle energy decreases by 2.5 percent.
The particle energy decreases because of deceleration of driving bunch
which then generates plasma wake (essentially relativistic Langmuir waves)
which is then absorbed by trailing bunch.
Although trailing bunch is accelerated overall particle energy decreases
as driving bunch is 4 times longer in x-direction. 
Accordingly we see from panel (b) shows
EM field energy normalized to its final simulation time
value increases, commensurately to the decrease of particle energy in panel (a).

\subsection{Run 4 --  acceleration in the GeV energy range, linear regime}

In Run 4 we keep everything as in Run 3 but we increased energy of driving and trailing bunches 
to 20.4 GeV also since the increase in energy allows for the driving bunch to move
longer distances without depletion, we let it run for ten times longer time interval, i.e. to 
$t \omega_{pe}=2000$. We also make simulation box ten times longer, while
keeping the same number of grid points. i.e. $c/\omega_{pe}$ is
resolved with also 24 grid points, which is a tolerable resolution.

Note that unlike in Run 3, where by time $t \omega_{pe}=200$
the driving bunch started to show signs of depletion,
in Run 4 by time $t \omega_{pe}=2000$ the bunch stays intact (see panels Fig.\ref{fig6}(d-f)) and in 
principle it would have been possible to continue the simulation
and accelerate the trailing bunch to even higher energies.

We gather from Fig.\ref{fig7}(c) that trailing bunch accelerated to 
approximately 21 GeV starting from 20.4 GeV.
This is because, as can be seen in Fig.\ref{fig6}(a-f), trailing bunch rides in the negative
wake-field of about $-  10^{10}$ V/m for ten times longer time compared to Run 3.

\subsection{Run 5 -- acceleration in the GeV energy range, blowout regime}

In Run 5 we keep everything as in Run 4 but now we employ blowout regime by increasing
driving and trailing bunch number densities, as stated in  table \ref{tab1}.
The motivation for considering this numerical run was to see whether trailing bunch
acceleration for possible in {\it 1D} and blowout regime at the same time.
It is well known that in the blowout regime of plasma wake-field acceleration, it is
mostly the transverse electric field that creates the density cavity (sometimes also called the bubble) 
behind the driving electron bunch. In 1D electrons cannot move in transverse direction that is why,
\citet{me2016}  concluded that they have not seen trailing bunch acceleration in their 1D simulations.
As will be shown below, this conclusion is only partially correct and placing the trailing bunch
in a suitable position makes its acceleration possible.
Here, we explored this topic further. Based on various runs to optimize
the acceleration, it was found that Run 5 provides favorable conditions.
Namely we had to increase distance between driving and trailing bunches 
from $10-4.5=5.5 c/\omega_{pe}$ in Runs 1 --4 to 
$94.5-4.5=90.0 c/\omega_{pe}$ in Run 5.
This was done in order to place the trailing bunch into middle of the negative electric field wake,
as can be seen in Fig.\ref{fig8}(a-f).
We also note from panel Fig.\ref{fig8}(c) that despite the fact driving bunch stays
intact (see panel Fig.\ref{fig8}(f)) the plasma wake becomes quite 
distorted by the end of simulation (see panel Fig.\ref{fig8}(c)).
This is quite different from Runs 1--4. At this stage it is unclear what is the source
of such distortion of $E_x(x, t \omega_{pe}= 2000) $.

Nonetheless, as can be seem from Fig.\ref{fig9}(c), starting from initial 20.4 GeV trailing bunch 
accelerates to 24 GeV.
So, it {\it is} possible to have plasma wake-field acceleration in 1D and blowout regime.
It was found that, optimally there should be approximately $(90-100) c/\omega_{pe}$
distance between trailing and driving electron bunches, because in 1D blowout regime
driving bunch's wake is much longer than in 3D. As in 1D case electrons
cannot move in the transverse direction, the wake becomes much longer compared to 3D (and 2D).

\subsection{Investigation of plasma temperature dependence}

In all numerical Runs 1--5 temperature of all plasma species
with set equal to $T=10^5$ K. 
Fluid-simulation results of \citet{bera16} were carried out for
cold beam-plasma system, while our numerical simulations
have finite temperature. In this subsection we aim
to investigate {\it whether} our conclusion that, an approximation for an
electron bunch not to evolve in time for few hundred plasma periods 
only applies when it is sufficiently relativistic,
i.e. $v_b=0.9999 c$ and not mere $v_b=0.99 c$, {\it depends
on temperature variation}.
In Fig.\ref{fig10} we present additional simulation results
as in Fig.\ref{fig1} but for numerical run similar to Run 2
now with 100 times hotter temperature and 10 times shorter domain length.
Because our grid length, as in every other PIC simulation,
is proportional to the Debye length, which in turn is proportional
to $v_{th,e}=\sqrt{k_B T/m_e}$ (i.e. $\lambda_D = v_{th,e}/\omega_{pe}$)
100 times hotter plasma requires 10 shorter domain length
otherwise end simulation time of $200 /\omega_{pe}$
would have to be altered. For clear comparison we wanted to keep
the same end simulation time.
We gather Fig.\ref{fig10} that results are not significantly different
from Fig.\ref{fig2}.

Next in Fig.\ref{fig11} we present additional simulation results
as in Fig.\ref{fig1} but for numerical run similar to Run 2
now with 100 times cooler temperature and 10 times longer domain length.
Again, we see in Fig.\ref{fig11} that results are not significantly different
from Fig.\ref{fig2}. Thus from both figures Fig.\ref{fig10} 
and Fig.\ref{fig11} our conclusion is not temperature
dependent.
The fact that in fluid simulation of \citet{bera16}
the beam is intact even for $v_b=0.99 c$
can be explained by the fact that they/fluid-approach
ignores wave-particle interactions.
Thus, in finite-temperature PIC simulation it is rather important to
have electron beam with speed very close to speed of light.

\section{conclusions}

Here we argue that in some laboratory, see e.g. Figure 1 from \citet{corde}, and probably 
most astrophysical situations such as jets in the vicinity
of black holes \cite{Ebisuzaki20149,
Ebisuzaki2014} and flares in solar atmosphere \cite{me2017}, 
plasma wake-field acceleration of electrons is one dimensional. 
Namely, variation transverse to the beam's motion can be ignored. Thus,
one dimensional (1D), particle-in-cell (PIC), fully electromagnetic simulations of 
electron plasma wake field acceleration were conducted in order to study the 
differences in electron plasma wake field acceleration in MeV versus GeV and
linear versus blowout regimes.
First, it has been shown that care needs to be taken when using
fluid simulations, as PIC simulations demonstrate that an approximation for an
electron bunch not to evolve in time for few hundred plasma periods 
only applies when it is sufficiently relativistic. Electron bunch
speed needs to be at least $v_b=0.9999 c$ and not just $v_b=0.99 c$.
We establish that injecting driving and trailing electron bunches
into plasmas with $n_0=5 \times 10^{22}$ m$^{-3}$,
produces electric fields of $- 10^{10}$ V/m, 
if the bunch density is one third as that of plasma (linear regime),
and $-  10^{11}$ V/m if the driving bunch density is $2.5 n_0$ (blowout regime).
We study the differences in the plasma wake created and
what is an optimal position of the trailing electron bunch.
Starting from initial 36 MeV trailing bunch with $n_b=0.3 n_0$
its acceleration to 85 MeV is easily possible
within 200 plasma periods. For greater times
the approximation for a driving
electron bunch not to evolve in time becomes invalid and driving bunch distorts.
Starting from initial 20 GeV trailing bunch with $n_b=0.3 n_0$
its acceleration to 21 GeV happens
within 2000 plasma periods. 
When we increase driving bunch density  to $n_b=2.5 n_0$,
starting from initial 20 GeV trailing bunch with $n_b=n_0$
its acceleration to 24 GeV occurs within 2000 plasma 
periods and plasma wake size is much larger, and therefore,
the distance between driving and trailing bunches must be commensurately increased.
Thus, it {\it is} possible to have plasma wake-field acceleration in {\it 1D} and blowout regime
at the same time. It was established that, optimally there should be
approximately $(90-100) c/\omega_{pe}$
distance between trailing and driving electron bunches, because in 1D blowout regime
driving bunch's wake is much longer than in 3D.

In summary, we show that in the linear regime and GeV energies,
the accelerating electric field generated by the
plasma wake is similar to the linear and MeV
regime. However, because GeV energy driving bunch
stays intact for much longer time, the final acceleration energies
are much larger in the GeV energies case.
In the GeV energy range and blowout regime
the wake's accelerating electric field is much
larger in amplitude compared to the linear case and also
plasma wake geometrical size is much larger.
Therefore, the correct positioning of the trailing bunch is important to
achieve the efficient acceleration of the trailing electron bunch.

%
%
%

\begin{acknowledgments}
This research utilized Queen Mary University of London's (QMUL) 
MidPlus computational facilities,       
supported by QMUL Research-IT.
\end{acknowledgments}


\end{document}